\documentclass[reprint,preprintnumbers,showpacs,amsmath,twocolumn,showkeys,aps,prl]{revtex4-1}
\pdfoutput=1
\usepackage{amssymb}
\usepackage{amsbsy}
\usepackage{amsmath}
\usepackage{graphicx}
\usepackage{dcolumn}
\usepackage{bm}
\usepackage{natbib}
\usepackage{epsfig}
\usepackage{epstopdf}
\usepackage{dsfont}

\usepackage{hyperref}
\usepackage[usenames,dvipsnames]{xcolor}
\hypersetup{colorlinks, linkcolor={red},citecolor={blue}, urlcolor={MidnightBlue}}

\usepackage{pdfpages}
\begin{document}

%\title{Dynamical Localization in a Disordered System: Suppressing the Relaxation of a Random Quantum Magnet by Coherent Periodic Drive}

\title{The fate of dynamical many-body localization in the presence of disorder}

\author{Analabha Roy$^1$, and Arnab Das$^2$}
\email{arnab.das.physics@gmail.com}
\affiliation{$^1$ Saha Institute of Nuclear Physics, $1$/AF Bidhannagar, Kolkata-$700064$ \\
$^2$ Indian Association for the Cultivation of Science, $2$A \& $2$B Raja S. C. Mullick Road,
Kolkata - $700032$}

\begin{abstract}
Dynamical localization is one of the most startling manifestations of quantum interference, 
where the evolution of a simple system is frozen out under a suitably tuned coherent periodic drive.   
Here, we show that, although any randomness in the interactions of a many body system kills dynamical localization eventually, spectacular remnants survive even when the disorder is strong. We consider a disordered quantum Ising chain where the transverse magnetization relaxes exponentially with time with a decay time-scale $\tau$ due to random longitudinal interactions between the spins. We show that, under external periodic drive, this relaxation slows down ($\tau$ shoots up) by {\it orders of magnitude} as the ratio of the drive frequency $\omega$ and amplitude $h_{0}$ tends to certain specific values (the freezing condition). If $\omega$ is increased while maintaining the ratio $h_0/\omega$ at a fixed freezing value, then $\tau$ diverges exponentially with $\omega.$ The results can be easily extended 
for a larger family of disordered fermionic and bosonic systems. 

\end{abstract}
\maketitle

The dynamics of quantum systems driven out of equilibrium by coherent periodic drives has
remained an intriguing topic of interest from the early days of quantum mechanics (see, e.g., \cite{Sakurai}) 
to date
\cite{Andre-1,Andre-2,AAR-PRL,AAR-Generic,Arimondo-Rev,Gauge-1a,Gauge-1b,Gauge-2,Manisha-Amit-Diptiman-da,
Prosen-1,Bastidas,victor,Sthitadhi,Analabha1,Analabha2,AD-KS-DS-BKC,Kris-Freezing,Ashhab-Nori,Luca-Rigol,Anushya,
Arimondo-1,Arimondo-2,Bloch-Periodic,Shengstock,DMF-IISER-Exp,Kaitzer,Gopar}. 
One interesting and well-known phenomenon in this field, where repeated quantum interference results in a
scenario which is quite counter-intuitive and unexpected from the classical point of view, 
is that of dynamical freezing in a quantum system under a periodic drive.  
Illustrious examples include the localization of a single quantum particle for all time while being forced periodically  
in free space (dynamical localization \cite{Dunlap}), or in one of the two wells of a double-well potential 
modulated sinusoidally (coherent destruction of tunneling \cite{Hanggi}). In both cases, this happens due to the coherent suppression of tunneling. 

A many-body version of this phenomenon, dubbed as dynamical many-body 
freezing, has also been observed both theoretically 
\cite{AD-DQH,SB-AD-SDG,AD-RM,Russomanno,Anatoli-Periodic,Sei-Book} 
and experimentally \cite{DMF-IISER-Exp}. Dynamical many-body freezing, however, is a 
more drastic version of dynamical localization: in the latter only the tunneling term is 
renormalized to zero by the external drive, resulting in 
localization in real space, while in the former the entire many-body Hamiltonian - with all its mutually
non-commuting terms - vanishes \cite{AD-DQH}. This implies 
freezing of any arbitrary initial state in the Hilbert space, rather than freezing of initial 
states localized in real space only. Intuitively, such an unequivocal freezing of all degrees of freedom 
of a many-body system seems possible only under very special circumstances, where certain simplicities in the structure of the Hamiltonian allow for such large-scale destructive quantum interference. Here, we demonstrate that such dynamical 
many-body freezing can have strong manifestations even in a system where dynamics 
is induced by interactions which are totally random.

The plan of the paper is as follows. After introducing the system and the drive, we briefly
sketch the content of our analytical approach to the problem. Then we discuss our main 
results in that light. The precise condition of maximal freezing is obtained from this
analysis. We also go beyond that, using exact numerical results for large system-sizes, averaged
over several disorder realizations, and discuss further characteristics of the phenomenon. 
Finally conclude with an outlook.
%%%%%%%%%%%%%%%%%%%%%%%%%%%%%%%%%%%%%%%%%%%%%%%%%%%%%%%%%%%%%%%%%%%%%%%%%%%%%%%%
%%%%%%%%%%%%%%%%%%%%%%%%%%%%%%%%%%%%%%%%%%%%%%%%%%%%%%%%%%%%%%%%%%%%%%%%%%%%%%%%
\noindent
%{\it The Model and the Phenomenon }
Consider the following disordered one-dimensional Ising chain, subjected to a 
sinusoidal transverse field: 
\begin{equation}
H(t) = -\alpha J \sum_{i}^{L-1} J_i \sigma^{x}_{i}\sigma^{x}_{i+1}  
-\sum_{i}^{L} \left\{h_{0}\sin{(\omega t)} + \alpha h_i\right\} \sigma^{z}_{i}, 
\label{H_OBC}
\end{equation}
\noindent where $\sigma^{\alpha}$'s ($\alpha = x,y,z$) are components of Pauli spins, $J_{i}$'s and 
$h_{i}$ are respectively the (quenched) interactions and on-site fields - both drawn randomly 
from a uniform distribution between $\left(-1,+1\right).$
The transverse field is subjected to an external  drive of frequency $\omega$ (period $T = 2\pi/\omega$) 
and amplitude $h_{0}$ (we set $\hbar = 1$).  
%%%%%%%%%%%%%%%%%%%%%%%%%%%%%%%%%%%%%%%%%%%%%%%%%%%%%%%%%%%%%%%%%%%%%%%%%%%%%%%%%%%%%%%%%%	
%Note: Change 'hires.eps' to 'lowres.pdf' in figs for arxiv submission
%%%%%%%%%%%%%%%%%%%%%%%%%%%%%%%%%%%%%%%%%%%%%%%%%%%%%%%%%%%%%%%%%%%%%%%%%%%%%%%%%%%%%%%%%%	
\begin{figure*}[t]
\ \epsfig{file=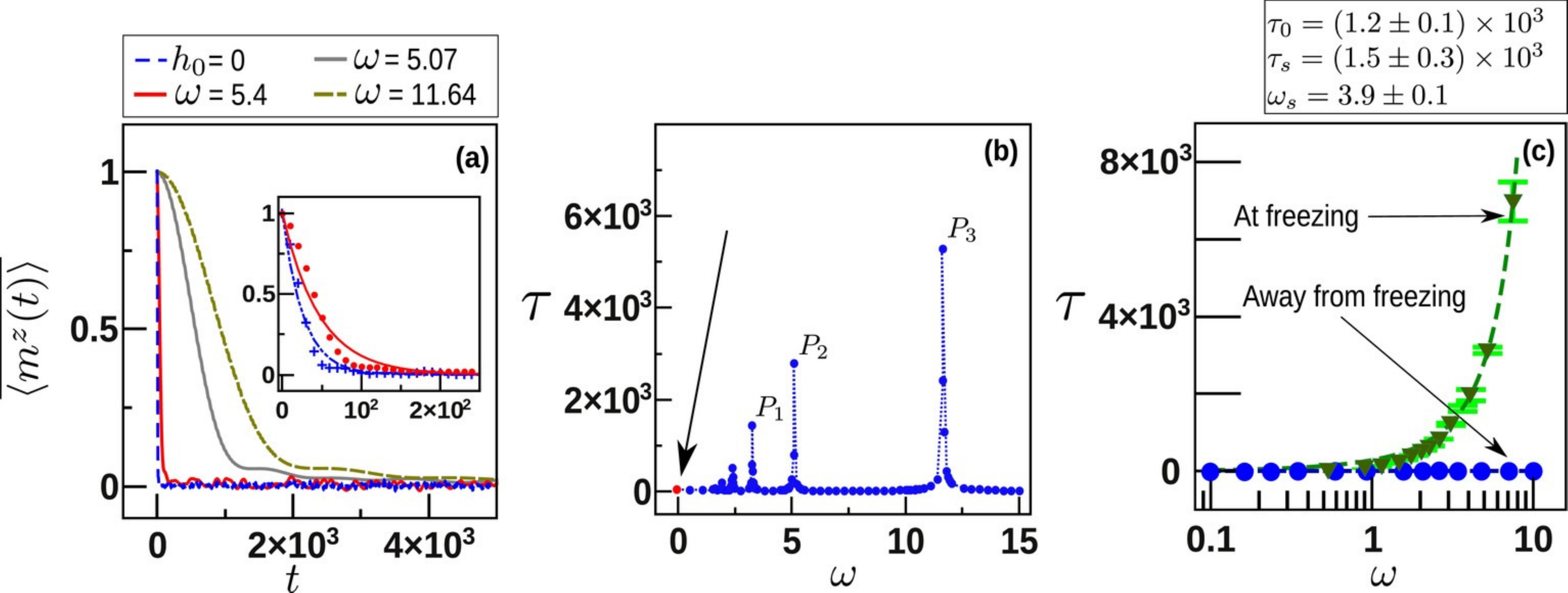,width=0.95\linewidth}
\caption{
(Color Online) Dynamical Freezing without field disorder ($h_i=0$):  
{\bf (a)} Exponential relaxation of the expectation value of ${m^{z}}$ with time for different
values of the drive frequencies $\omega$. Unless otherwise indicated, the drive amplitude $h_{0}$ is fixed at $7.0$. 
For certain specific values of $\omega$ (e.g., $\omega=5.07,11.64$), the relaxation is 
tremendously slowed down. The relaxation in the 
absence of the drive is labeled separately for comparison. The inset compares a representative sample of the numerical data (shown as points) to the curves fitted to them using Eq.~\ref{Exp-decay-fit}.  
{\bf(b)} {\it $\tau$ vs $\omega$ for fixed $h_{0}$}: 
The sharp peaks indicate dramatic enhancements of $\tau$ for certain values of $\omega$. 
Three of the most prominent peaks are identified
as $P_{1-3}$. The values of $\omega$ at these peaks are $P_1\approx3.24$, $P_2\approx5.07$, 
and $P_3\approx11.64$. Those 
values are identified to be the ones for which the effective Hamiltonian $H_{\rm eff}$
(Eq.~\ref{eq:heff}) vanishes. The red dot pointed by the arrow-head represents the 
case in absence of the drive. 
{\bf (c)} {\it $\tau$ vs $\omega$ at fixed $\eta$}:
Comparison of enhancement of $\tau$ as $\omega$ is increased keeping $\eta = \frac{4h_{0}}{\omega}$ fixed 
for two cases - under the freezing condition ${\mathcal J}_{0}(\eta) = 0$ ($\eta \approx 2.4048$), and away from it 
${\mathcal J}_{0}(\eta) \approx 0.765$
($\eta = 1.0$) as marked in the Fig. 
Exponential enhancement of $\tau$ with $\omega$ is observed (numerical data fitted with the
$\tau(\omega)\bigr|_{{\mathcal J}_{0}(\eta)=0} =  \tau_{0} + \tau_{s}e^{{\omega}/{\omega_{s}}}$
form)  under the freezing condition, while no noticeable variation of $\tau$ is observed away from the
freezing condition. 
Results are for $L=100,$ averaged over $> 10^3$ 
disorder realizations of the bonds $J_i$.
The error-bars due to disorder-induced fluctuations are about the point size (see e.g., Fig. S1 of Suppl. Matt.),
and hence omitted (except for (c)) to avoid cluttering.}  
\label{mz-Rlx-tau}
\end{figure*}

\noindent
The Hamiltonian in Eq.~\ref{H_OBC}
can be mapped to the Hamiltonian~\cite{Lieb,Young-Random,Dziarmaga-Random},
\begin{multline}
H(t) = - \alpha J \sum_{i}^{L}  J_i \left(c^{\dagger}_i c^{\dagger}_{i+1} 
+ c^{\dagger}_i c_{i+1}  + {\rm h.c.}\right) \\
- 2 \sum_i^{L} \left\{h(t)+\alpha h_i\right\} c^{\dagger}_i c_i,
\label{H_Fermi}
\end{multline}
\noindent
with hard-core bosons  created (annihilated) by $c^\dagger_i$ ($c^{\;}_i$)
These bosons satisfy $\{ c_{j}^{\dagger},c_{j} \}=0$, and the usual bosonic commutation relations for $i\ne j$. Also, $h(t)=h_0\sin{\omega t}$.

In order to gain insight into the drive-dependent sharp jumps in the relaxation time-scale,
we follow a recently developed scheme \cite{Mintert} of deducing a renormalized time-independent effective 
Hamiltonian $H_{\rm eff}$ which describes the evolution of the system if observed stroboscopically at instants
$t = nT$, where $n$ denotes natural numbers: $U(nT) = e^{-iH_{\rm eff}nT},$ 
where $U(t) = {\mathcal T}\exp{[-i\int_{0}^{t}H(t^{\prime})dt^{\prime}]}$ (${\mathcal T}$ denotes time-ordering).  
For $\omega \gg |\alpha J|$, \textit{i.e.} in the limit of fast drive, no appreciable evolution takes place within a 
period, and stroboscopic observations represent smooth evolutions to a good approximation.  
It follows from Floquet theory~\cite{Floquet-Original} that for a $T$-periodic Hamiltonian $H(t),$ 
the time-evolution operator can be written as $U(t) = e^{-iH_{\rm eff}t}Z(t),$ where $Z(t)$ is $T$-periodic 
and $H_{\rm eff}$ is Hermitian. Clearly, $H_{\rm eff}$ is an operator that is largely non-local in the original 
degrees of freedom and is not necessarily amenable to any simple physical interpretation. 
However, since $Z(0)=Z(nT)=\mathds{1}$ (identity), $H_{\rm eff}$ can be considered as
an effective time-independent Hamiltonian giving the correct stroboscopic evolution. 
Using the above decomposition of $U(t)$ and the time-dependent Schr\"{o}dinger equation it satisfies, we get
\begin{equation} 
H_{\rm eff} = Z^{\dagger}(t)H(t)Z(t) -iZ^{\dagger}(t)(\partial_{t}Z(t)).
\label{Heff} 
\end{equation}
\noindent
Deducing the exact form of $H_{\rm eff}$ from the above equation is as hard as solving the original 
time-dependent problem. However, controlled approximations in the large $\omega$ limit can
be obtained \cite{Mintert} using a flow equation technique \cite{Wegner,Kehrein-book} 
which we employ here. This gives (see Suppl. Matt. for details)
\begin{multline}
\label{eq:heff}
H_{\rm eff}    =   -J\sum_i j^{(0)}_i\left(c^\dagger_i c^\dagger_{i+1} + \rm{h.c.}\right) \\
		            -J\sum_i j^{(1)}_i\left(c^\dagger_i c^{\;}_{i+1}+\rm{h.c}\right) 
		            -\mu\sum_i j^{(2)}_i c^\dagger_i c^{\;}_i,
\end{multline}
with $\eta \equiv 4h_0/\omega$, and the constants $j^{(s)}_i, \mu $ defined as follows.
\begin{eqnarray}
j^{(0)}_i            &\equiv& \alpha J_i\; \bigg\{\mathcal{J}_0(\eta) - \frac{4\alpha h_i}{\omega}\beta(\eta)\bigg\} ,\nonumber \\
j^{(1)}_i            &\equiv& \alpha J_i \; \mathcal{J}_0(\eta) ,\nonumber \\
j^{(2)}_i            &\equiv& \frac{h_i}{J},\nonumber \\
\mu 		     &\equiv& 2\alpha J.
\label{Renorm_fctr}
\end{eqnarray}
Here, $\mathcal{J}_0(\eta)$ is the ordinary Bessel function of order zero
(note that the way they are defined in Eq.~(\ref{H_OBC}), $J_{i}$s are 
dimensionless and $h_{i}$'s have dimension of energy). In addition, the function $\beta(\eta)\equiv\sum_{n\neq 0} {\mathcal{J}_n(\eta)}/{n}$. 

The above expression for $H_{\rm eff}$ is obtained under a rotating-wave approximation
(RWA) which holds for $\omega \gg J,\alpha.$ This effective Hamiltonian accurately reproduces the dynamics 
of the full system stroboscopically to the lowest order in $\alpha/\omega$~\cite{suppl}.
%%%%%%%%%%%%%%%%%%%%%%%%%%%%%%%%%%%%%%%%%%%%%%%%%%%%%%%%%%%%%%%%%%%%%%%%%%%%%%%%%%%%%%%%%%	
%Note: Change 'hires.eps' to 'lowres.pdf' in figs for arxiv submission
%%%%%%%%%%%%%%%%%%%%%%%%%%%%%%%%%%%%%%%%%%%%%%%%%%%%%%%%%%%%%%%%%%%%%%%%%%%%%%%%%%%%%%%%%%
\begin{figure}[htb]
\ \epsfig{file=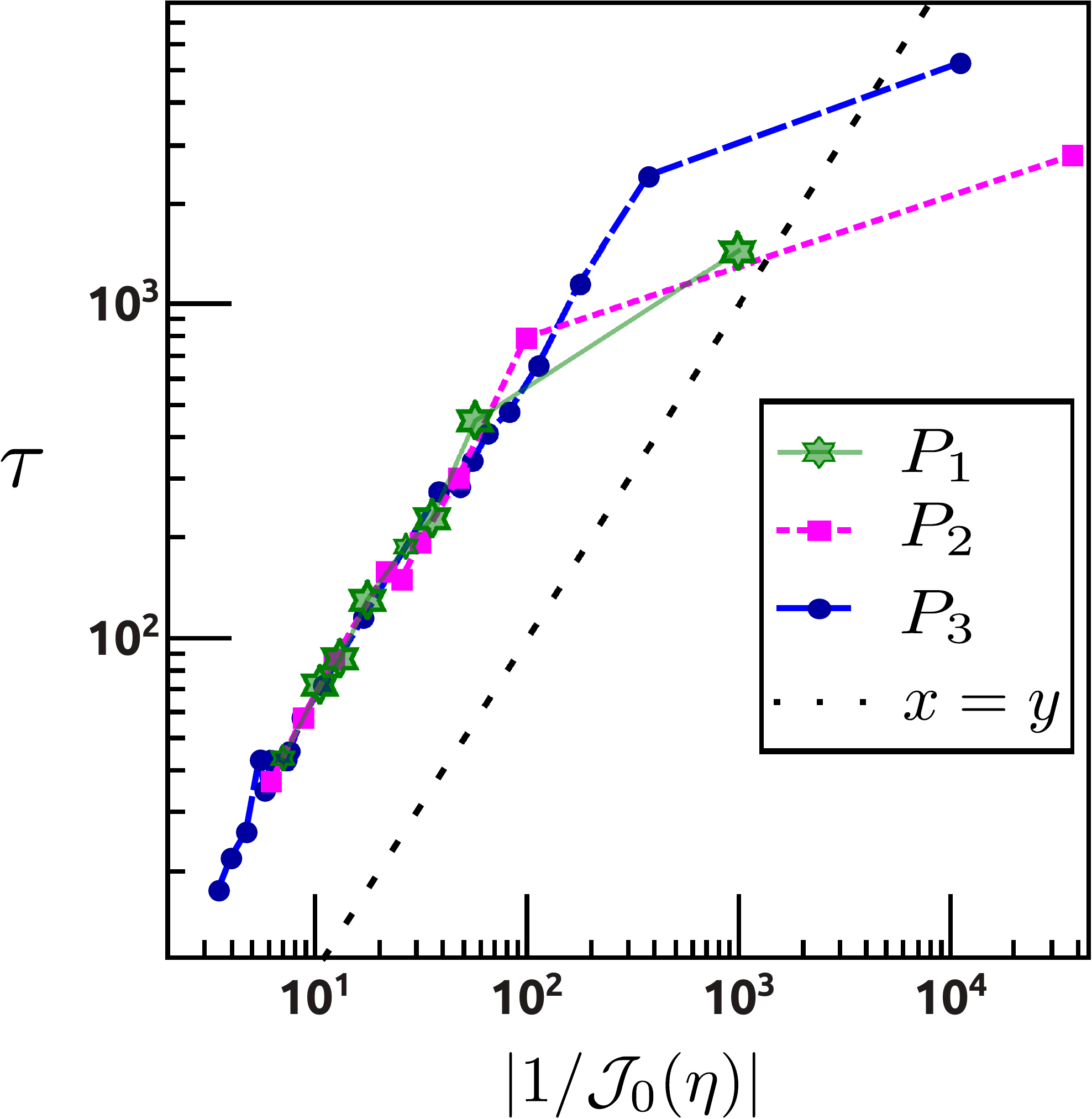,width=0.7\linewidth}
\caption{(Color Online) Dependence of $\tau$ on the drive parameters.
Numerical data for $\tau$ vs $1/{\mathcal J}_{0}(\eta)$ for $h_{i}=0$, $h_{0}=7$ 
(for a single random realization) are shown for the range of $\omega$ around the peaks $P_{1,2,3}$ 
of Fig.~\ref{mz-Rlx-tau}(b). The black dashed line is a guide to the eye. 
As expected from Eqs.~(\ref{eq:heff}, \ref{Renorm_fctr}),
we observe $\tau \propto 1/{\mathcal J}_{0}(\eta)$ where ${\mathcal J}_{0}(\eta) \gg  \alpha/\omega^{2}$.
However, for small ${\mathcal J}_{0}(\eta),$ $\tau$ tends to saturate to a large value \textit{viz.} the respective 
peak values in Fig.~\ref{mz-Rlx-tau}(b) (instead of diverging).}
\label{tau-vs-J0}
\end{figure}
%%%%%%%%%%%%%%%%%%%%%%%%%%%%%%%%%%%%%%%%%%%%%%%%%%%%%%%%%%%%%%%%%%%%%%%%%%%%%%%%%%%
Now we describe our main results for the driven spin Hamiltonian $H(t)$ 
and analyze it in light of the effective time-independent Hamiltonian $H_{\rm eff}$.
To avoid cluttering, the results presented in Fig.~\ref{mz-Rlx-tau} and the main discussion
are focused on the case where the field disorder is absent ($h_{i}=0$), the case with $h_{i}\ne 0$ is given afterwards.
Imagine that the spins (in $H(t)$) are initially prepared in a state strongly polarized in $+z$-direction 
and the transverse magnetization $ m^{z} = \frac{1}{L}\langle\sum_{i=1}^{L}\sigma_{i}^{z}\rangle \approx 1$. 
(here $\langle ... \rangle$ denotes quantum expectation values, 
and $\overline{{\mathcal O}}$ denotes average of ${\mathcal O}$ over disorder realizations).
In the absence of the drive ($h_{0}=0$), the magnetization $m^z$ relaxes with time since the random interaction terms in the Hamiltonian do not commute with it. If the drive is switched on, the characteristic time-scale of the relaxation 
changes, depending on the drive. The relaxations of $\overline{m^{z}}$ with time for various drive 
frequencies are shown in Fig.~\ref{mz-Rlx-tau}(a) (these results are obtained by numerically solving the 
time-dependent Schr\"{o}dinger equation for several disorder realizations and averaging 
over them). The relaxations (both in the absence and presence of the drive) 
are fitted accurately with the exponential decay form 
(inset of Fig.~\ref{mz-Rlx-tau}(a)).
\begin{equation}
\overline{\langle m^{z}(t)\rangle} =  m^{z}_{0}e^{-t/\tau},
\label{Exp-decay-fit}
\end{equation} 
\noindent
where $\tau$ is the decay constant. Our main result concerns the behavior of the decay time-scale $\tau$ 
as a function of the drive frequency $\omega$ for a given drive amplitude $h_{0}$. \\

\noindent
{\it Freezing Points:}
The behavior is quite dramatic, as shown in Fig.~\ref{mz-Rlx-tau}(b):
for certain values of $\omega$, the value of $\tau$ shoots up sharply 
by orders of magnitude compared to the undriven case. This corresponds 
to the extreme slowing down (freezing) of the decay of $m^z$ visible for certain 
values of $\omega$ in Fig.~\ref{mz-Rlx-tau}(a).
This can be explained by noting that for $h_{i}=0$ (as considered in the figure), 
the effective Hamiltonian  $H_{\rm eff} \propto {\mathcal J}_{0}(\eta)$ to leading order in $\alpha/\omega$ 
(Eq.~\ref{eq:heff},\ref{Renorm_fctr}) and hence $\tau \propto 1/{\mathcal J}_{0}(\eta)$ for $\omega \gg J,\alpha,$ (see Fig.~\ref{tau-vs-J0}). Thus, there is an overall renormalization of the time-scale, which can be controlled
by the factor ${\mathcal J}_{0}(\eta).$ Under the special condition ${\mathcal J}_{0}(\eta)=0$ (the freezing condition), $H_{\rm eff}$ 
vanishes, indicating a large enhancement of {\it all} timescales observable in the dynamics governed by $H_{\rm eff}$.
Interestingly, this implies that the slowing down is not only limited to $m^z(t)$, 
or to any special initial condition. 
For finite $\omega$, the approximation is, of course, valid only as long as 
${\mathcal J}_{0}(\eta) \gg \alpha/\omega^{2}$ -- otherwise higher order terms in $\alpha/\omega$ 
come into play. This results in a saturation due to the breakdown in the linear behaviour of $\tau$ with $1/{\mathcal J}_{0}(\eta)$ (Fig.~\ref{tau-vs-J0}). The saturation is also seen in the large but finite values of $\tau$ observed
at the freezing points ($P_{1,2,3}$) of Fig.~\ref{mz-Rlx-tau} (b), instead of infinite $\tau$ as suggested by 
the vanishing of $H_{\rm eff}$ at those points.    \\

\noindent{\it Beyond the Rotating-Wave Approximation:}
Though $\tau$ does not diverge for finite $\omega$ due to higher order corrections to 
RWA, those corrections should vanish as $\omega \to \infty,$ 
resulting in $\lim_{\omega\to\infty}\tau\to\infty$
under the freezing condition ${\mathcal J}_{0}(\eta)=0.$ To characterize the dependence of $\tau$ on $\omega$ 
under the freezing condition going beyond RWA, 
we numerically study the variation of $\tau$ with $\omega$ fixing $\eta$ to a freezing value. 
Fig.~\ref{mz-Rlx-tau} (c) shows that $\tau$  
diverges exponentially with $\omega$. The numerical results in the figure are  
fitted well with the form $\tau(\omega)\bigr|_{{\mathcal J}_{0}(\eta)=0} =  \tau_{s}e^{{\omega}/{\omega_{s}}}-\tau_0$. 
If $\eta$ is held fixed to some value such that ${\mathcal J}_{0}(\eta) \not\approx 0$ (away from freezing), $\tau$ does not show
any appreciable change with increase in $\omega$ within the range considered (Fig.~\ref{mz-Rlx-tau} c). In this range,
$\tau$ increased by orders of magnitude for the freezing case. \\

\noindent{\it The $\omega \to \infty$ limit:} 
Note that absolute 
freezing, \textit{i.e.} the divergence of $\tau$ in the $\omega \to \infty$ limit 
under the freezing condition, is a counter-intuitive result. Intuitively, an infinitely 
fast purely sinusoidal drive should simply be equivalent to the absence of the drive altogether, 
since the driven parameters return back to themselves within no appreciable time in each cycle. 
This would mean that $m^z$ should decay when $\omega \to \infty$ 
as if there was no drive at all. This is indeed the case away from freezing -- in the large $\omega$
range we considered, (see Fig~\ref{mz-Rlx-tau} c), $\tau$ remains same 
as that of the undriven case as $\omega$ is increased. However, 
under the freezing condition, $\tau$ diverges exponentially, indicating that the decay will completely
stop due to the drive as $\omega \to \infty$. \\
%%%%%%%%%%%%%%%%%%%%%%%%%%%%%%%%%%%%%%%%%%%%%%%%%%%%%%%%%%%%%%%%%%%%%%%%%%%%%%%%%%%%%%%%%%	
%Note: Change 'hires.eps' to 'lowres.pdf' in figs for arxiv submission
%%%%%%%%%%%%%%%%%%%%%%%%%%%%%%%%%%%%%%%%%%%%%%%%%%%%%%%%%%%%%%%%%%%%%%%%%%%%%%%%%%%%%%%%%%
\begin{figure*}[t]
\ \epsfig{file=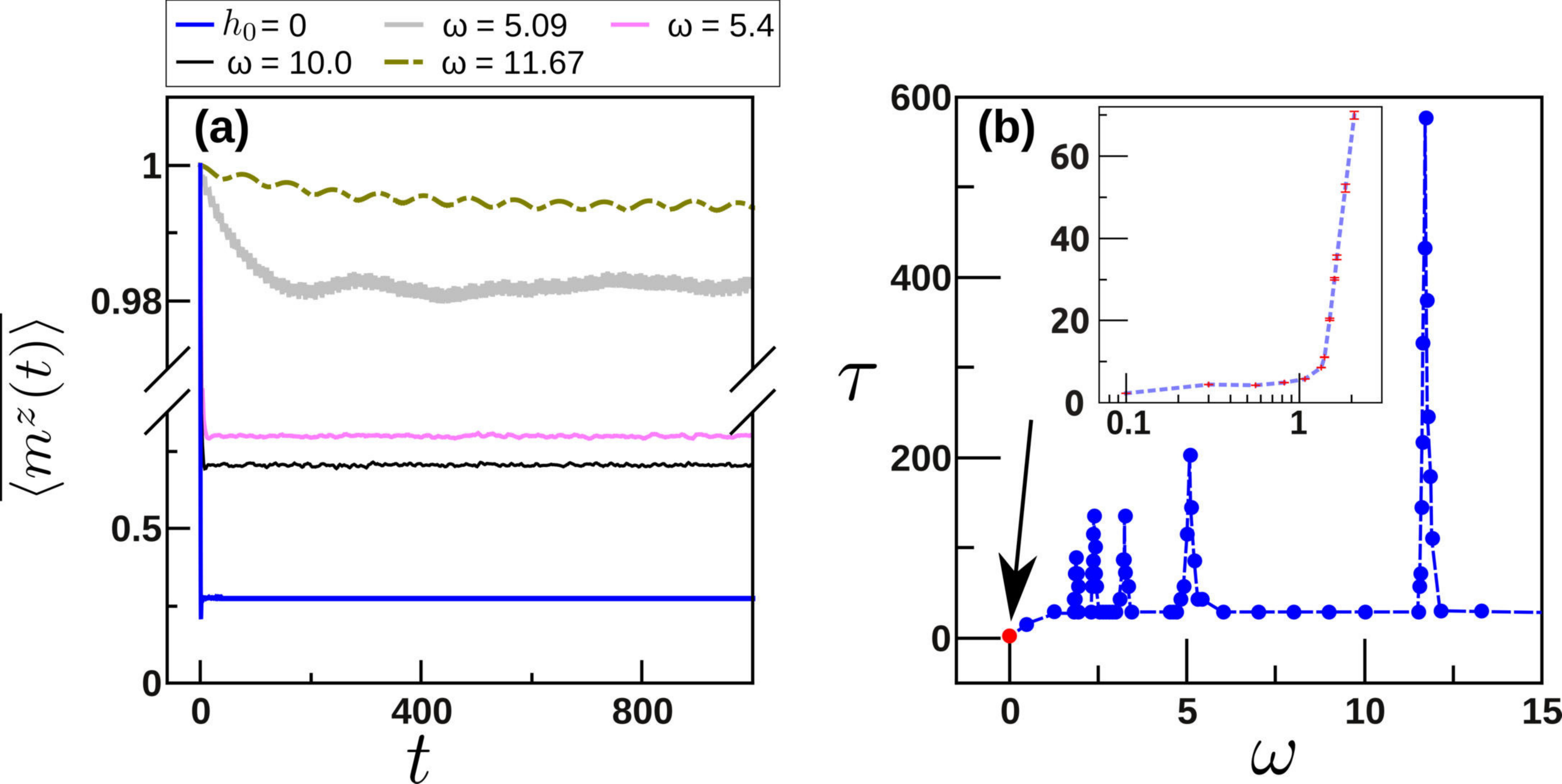,width=0.7\linewidth}
\caption{
(Color Online) Dynamical Freezing with field disorder:  
{\bf (a)} Exponential relaxation of $m^{z}$ with time for different
values of the drive frequencies $\omega$. {\bf (b)} 
The timescale $\tau$ of the exponential relaxation of $m^z$ vs $\omega$. 
Note the peaks at the roots of $\mathcal{J}_0(\eta)$, the same places as peaks 
$P_{1-3}$ in Fig~\ref{mz-Rlx-tau}. The red dot pointed by the arrow-head represents 
the case in absence of the drive. The inset plots $\tau$ vs $\omega$ (log-scale) at 
smaller values while adjusting $h_0$ so as to remain at the first root of  $\mathcal{J}_0(\eta)$. 
At very small values of $\omega$, the higher order contributions destroy the dynamical 
equivalence between $H(t)$ and $H_{\rm eff}$, leading to a loss of freezing. 
Freezing begins to appear as $\omega$ is increased, as seen in the inset by the growth of $\tau$ with $\omega$
for $\eta$ fixed to a freezing value. All the parameters are the same 
as in Fig~\ref{mz-Rlx-tau}, except with field disorder $h_i$ switched on.
$h_{i}$ is distributed uniformly and randomly between $\pm 1$}. 
\label{mz-Rlx-tau-fdon}
\end{figure*}
%%%%%%%%%%%%%%%%%%%%%%%%%%%%%%%%%%%%%%%%%%%%%%%%%%%%%%%%

\noindent
{\it Static disordered field ($h_i\neq 0$):}
Now we briefly present the results for the case with quenched disorder in the transverse field.
Including randomness in the field does not alter the basic features of the phenomenon discussed above, 
but there are some quantitative differences that we discuss here. In this case $H_{\rm eff}$
does not vanish even to first order in $1/\omega$ at the freezing point due to the field-dependent 
terms in Eq.~\ref{eq:heff} that scale inversely with $\omega$. Thus, we expect freezing at the roots of 
$\mathcal{J}_0(\eta)$ to last for shorter times at large $\omega$ when the field 
disorder is on. This is qualitatively verified by our numerical simulations of the 
exact dynamics with field disorder on. Figure~\ref{mz-Rlx-tau-fdon} shows plots of $m^z$ and the 
exponential relaxation time scale therein in regimes similar to Fig.~\ref{mz-Rlx-tau}, except with 
the field disorder switched on. The plots show that freezing is maintained at the roots of 
$\mathcal{J}_0(\eta)$, although the time scale of the decay is one order of magnitude less than the 
case without any static field.

\noindent
{\it Summary and Outlook:}
We have shown that dynamical localization can have drastic manifestations in many-body systems with
extensive disorder. We show that the application of a coherent periodic drive with specific values of the ratio 
of the drive frequency $\omega$ and the amplitude $h_{0}$ (freezing condition), can drastically slow 
down the natural relaxation induced by random interactions between the spins in a disordered Ising chain
for {\it any arbitrary initial state in the Hilbert space}. 
For moderately high values of $\omega$ and $h_{0}$, $\tau$ is observed to be orders of magnitude higher 
than its value away from freezing (or that in absence of the drive). At a specific freezing point 
($h_{0}/\omega$ kept fixed), the characteristic relaxation time $\tau$ diverges exponentially with 
$\omega$. However, if the ratio is fixed away from the freezing value, increasing $\omega$
does not have any observable effect on $\tau.$ These results are the first of their kind, showing drastic survival of 
dynamical localization, which is the result of large-scale destructive quantum interference induced by a 
periodic drive, in a highly disordered system. This opens up possibilities of preserving arbitrary (even unknown) 
quantum states of interacting qubits (Ising spins) from decaying due to unknown stray interactions.   
Experimental realizations of bosons in disordered 1d potentials in optical lattices have already been achieved 
experimentally~\cite{Disordered-Bosons,Bloch-Nat-Phys}. Coherent periodic drives applied to hardcore bosons
in optical lattices, used for realizing spin chains with precise control over the Ising-like
couplings, have also materialized~\cite{Bloch-Periodic}.
Hence, experimental observations of the freezing phenomenon seem feasible 
in a present day cold-atom laboratory.       \\
\noindent
{\it Acknowledgements:}
AR thanks CSIR, India for support under Scientists' Pool Scheme No. $13(8531-$A$)/2011/$Pool, as well as the TACC, University 
of Texas at Austin, for access to their clusters. Both authors thank Kasturi Basu, IACS Kolkata, for useful discussions.

\bibliographystyle{apsrev4-1}
\bibliography{bib_freezing} 
 
%%%%%%%%%%%%%%%%%%%%%%%%%%%%%%%%%%%%%%%%%%%%%%%%%%%%%%%%%%%%%%%%%%%%%%%%%%%%%%%%%%%%%%%%%%	
%Note: Uncomment these for arxiv submission. Update PDF filename and add pages if needed
%%%%%%%%%%%%%%%%%%%%%%%%%%%%%%%%%%%%%%%%%%%%%%%%%%%%%%%%%%%%%%%%%%%%%%%%%%%%%%%%%%%%%%%%%%
\clearpage 

\section*{Supplementary material}

# Supplementary Materials

In these supplementary notes, we detail some of the methodologies used to obtain the results detailed in the main paper. Our starting point is the transformed Hamiltonian in Eq. (2) of the main paper *viz.*

$$H(t) = -\alpha J \sum_i^L J_i \left( c_i^\dagger c_{i+1}^\dagger + c_i^\dagger c_{i+1} + \text{h.c.} \right) - 2 \sum_i^L \{h(t) + \alpha h_i\} c_i^\dagger c_i, \tag{S1}$$

with $h(t) = h_0 \sin \omega t$. The first section elaborates on the numerical methods used to gather the time-magnetization data from the Hamiltonian above. The second section details the Floquet-RG flow equations used to pinpoint the approximate unitary transformation needed to obtain the effective Hamiltonian $H_{\text{eff}}$ in the main paper. The final section details comparison of numerical evolution of the $H_{\text{eff}}$- dynamics with that of the full Hamiltonian, as well as key qualitative observations in the former observed in the latter.

## NUMERICAL METHODOLOGY:

In order to numerically simulate the time-dynamics of the Hamiltonian in Eq. S1, we rewrite it in terms of free fermions [r1]. Though the transformation is non-local in nature, we exploit a natural ordering in the sites to yield a Hamiltonian identical to Eq. S1 (except with free fermions instead of hard-sphere bosons) by canceling out nonlocal contributions. The present disordered case can be block-diagonalized by rewriting Eq. S1, with free fermions, in terms of Nambu spinors $\Psi$:

$$H(t) = \Psi^\dagger \bar{H}(t) \Psi,$$
$$\Psi^\dagger \equiv \begin{pmatrix} c_1^\dagger & c_2^\dagger & \ldots & c_L^\dagger, c_1 & c_2 & \ldots & c_L \end{pmatrix},$$
$$\bar{H}(t) \equiv \begin{pmatrix} A(t) & B(t) \\ -B(t) & -A(t) \end{pmatrix}. \tag{S2}$$

Here, $A$ and $B$ are real $L \times L$ matrices. The matrix $A = A^d(t) + A^o$, where $A_{i,j}^d(t) = -[h(t) + \alpha h_i]\delta_{i,j}$ is diagonal. The non-zero elements of $A^o$ and $B(t)$ are given by $A_{i,i+1}^o = A_{i+1,i}^o = -\alpha J J_i/2$, $B_{i,i+1}(t) = -B_{i+1,i}(t) = -\alpha J J_i/2$. The Heisenberg equations of motion for the operator $c_{i,H}(t)$ are

$$i\frac{d}{dt} c_{i,H}(t) = U^\dagger(t) [c_i, H(t)] U(t), \tag{S3}$$

where the subscript $H$ denotes operators in the Heisenberg picture, and $U(t)$ is the time translation that solves the Schrödinger equation of the Hamiltonian in Eq. S1. Note that, for every operator $O(t)$ in the Schrödinger picture, the corresponding operator in the Heisenberg picture, $O_H(t) = U^\dagger(t) O(t) U(t)$. We now substitute the last of Eq. S2 into the rhs of Eq. S3 to yield

$$i\frac{d}{dt} c_{i,H}(t) = 2 \sum_j \left[ A_{ij}(t) c_{j,H}(t) + B_{ij}(t) c_{i,H}^\dagger(t) \right]. \tag{S4}$$

These equations can be solved by the trial solution $c_{i,H}(t) = \sum_\mu \left[ u_{i\mu}(t) \gamma_{\mu,H}(t) + v_{i\mu}^*(t) \gamma_{\mu,H}^\dagger(t) \right]$, where the Bogoliubov ansatz makes the many body state $|\Psi(t)\rangle$ the emergent vacuum of the modes created by $\gamma_\mu^\dagger(t)$ $\forall t$ *viz.* $|\Psi(t)\rangle \sim \prod_\mu \gamma_\mu(t)|0\rangle$, where the normalization has been suppressed, and the product is over all lattice sites. A necessary and sufficient condition for the emergent modes to generate the emergent vacuum is that $\gamma_\nu(t)|\Psi(t)\rangle = 0$, leading to $\frac{d}{dt}(\gamma_\nu(t)|\Psi(t)\rangle) = 0$. In the Heisenberg picture, this condition can be rewritten as $i\frac{d}{dt}\gamma_{\mu,H}(t) = 0$. Using this result, we substitute the trial solution into Eq. S4 and compare operator coefficients. This leads to the following dynamical system:

$$i\frac{d}{dt} u_{i\mu}(t) = 2 \sum_{j=1}^L \left[ A_{i,j}(t) u_{j\mu}(t) + B_{i,j}(t) v_{j\mu}(t) \right]$$
$$i\frac{d}{dt} v_{i\mu}(t) = -2 \sum_{j=1}^L \left[ A_{i,j}(t) v_{j\mu}(t) + B_{i,j}(t) u_{j\mu}(t) \right]. \tag{S5}$$

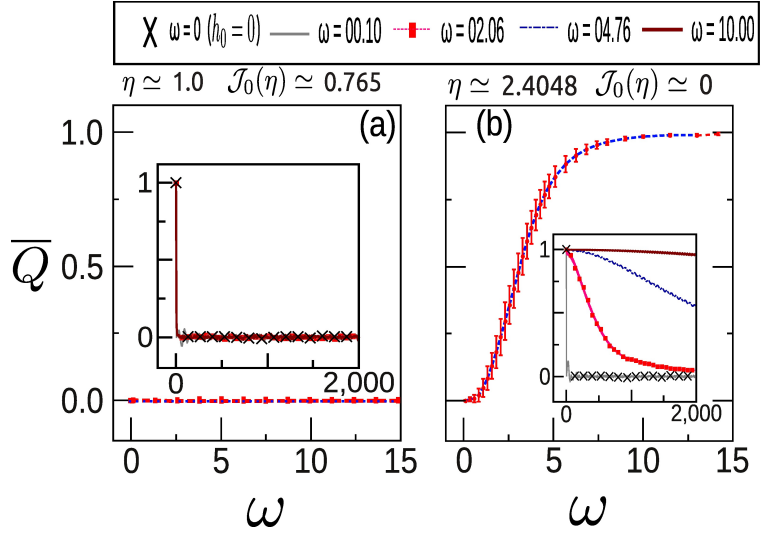

FIG. S1. (Color Online) **Disorder fluctuations in the numerics: Nonfreezing (a) vs Freezing (b)**: (a) Disorder average of magnetization (*i.e.* $\overline{Q}$ where $Q \equiv \langle m^z(t) \rangle$), as well as fluctuations (*i.e.* $(\overline{Q^2} - \overline{Q}^2)^{1/2}$) as a function of $\omega$ (main-frame) with $h_0$ adjusted to keep $\eta \equiv 4h_0/\omega = 1.0$, away from the freezing condition *i.e.* $\mathcal{J}_0(\eta) \neq 0$; **inset** shows $\overline{m^z}$ vs $t$ - for corresponding cases. (b) Under the freezing condition, $\mathcal{J}_0(\eta) \approx 0$ at the first zero of the Bessel function. Here, $\overline{Q}$ approaches unity with increasing $\omega$ as predicted by the theory - the effective Hamiltonian (Eqs. S23) vanishes to order higher than one in $1/\omega$. In all plots, the field disorder is absent ($h_i = 0$), $\hbar = 1$, $J = 1$, and the disorder strength $\alpha = 0.3$. The red error bars represent disorder fluctuations, and are negligible away from freezing (panel(a)), as well as at freezing for large $\omega$. The largest fluctuations occur when the freezing condition breaks down due to the contributions of higher order terms in the effective Hamiltonian, but do not go above 0.14 as seen in panel (b).

The objects $u_{j\mu}(t)$, $v_{j\mu}(t)$ can be encapsulated into $L \times L$ matrices $u(t)$ and $v(t)$. Unitary evolution and canonicality demands that $\{\gamma_\mu(t), \gamma_\nu^\dagger(t)\} = \delta_{\mu\nu}$, $\{\gamma_\mu(t), \gamma_\nu(t)\} = \{\gamma_\mu^\dagger(t), \gamma_\nu^\dagger(t)\} = 0$. Thus, $u^\dagger(t)u(t) + v^\dagger(t)v(t) = \mathbb{1}$, and $v^T(t)u(t) + u^T(t)v(t) = 0$. Any initial condition we choose must satisfy these conditions at $t = 0$. Our choice of initial condition *viz.* the classical ground state strongly polarized in the $+z$-direction, translates to $|\Psi(0)\rangle \sim \prod_i c_i^\dagger |0\rangle$. This can be set by putting $u(0) = 0$, $v(0) = \mathbb{1}$.

The magnetization per site can be shown via Jordan-Wigner transformation to be $\langle m^z(t) \rangle \equiv -1 + \frac{2}{L} \sum_i \langle \Psi(0) | c_{i,H}^\dagger(t) c_{i,H}(t) | \Psi(0) \rangle$. Substituting the trial solution and using the canonical anti-commutation relations of $\gamma_\mu(t)$ simplifies the magnetization to

$$\langle m^z(t) \rangle = -1 + \frac{2}{L} \operatorname{Tr}\left[ v^\dagger(t) v(t) \right]. \tag{S6}$$

The following exponential form fits the exact numerical data for the disorder-averaged magnetization very well -

$$\overline{\langle m^z(t) \rangle} = m_\infty^z + (m_0^z - m_\infty^z) e^{-t/\tau}, \tag{S7}$$

where $m_0^z$ ($\approx 1$ for our case) and $m_\infty^z$ are the values of $m^z$ at the initial and the infinite time limit respectively, and $\tau$ is the decay time-scale. Here, just as in the main paper, $\langle \ldots \rangle$ denotes quantum expectation values, and $\overline{\mathcal{O}}$ denotes average of operator $\mathcal{O}$ over disorder realizations. If we turn off the field disorder *i.e.* set $h_i$ in Eq. S1 to 0 $\forall i$, then $m_\infty^z \approx 0$, yielding Eq. (2) in the main paper.

All the figures in the main paper are obtained by solving Eqns. S5 numerically using the Runge-Kutta Prince-Dormand (8, 9) routine from the GNU Scientific Library [r2], and matrix operations (such as the evaluation of Eq. S6) were vectorized using its associated BLAS library [r3]. Multiple disorder realizations for the same parameter set were built by using the luxury random number generator in the GNU Scientific Library [r2], and were run in parallel using the Message Passing Interface. The disorder averages of time-averaged magnetization, denoted by $\overline{Q}$ where $Q \equiv \langle m^z(t) \rangle$, have already been plotted in the main paper. The disorder fluctuations were typically quite small for a large range of drive parameters at $\alpha = 0.3$, and were omitted in the main paper. For the sake of completeness, the disorder fluctuations in the time-averaged magnetization are shown for a representative range of drive parameters in Fig. S1 .

2

## ANALYTICAL TREATMENT VIA UNITARY RENORMALIZATION GROUP FLOW IN FLOQUET SPACE

In this section, we elaborate on our analytical treatment of this problem by mapping the $T = 2\pi/\omega$ - periodic Hamiltonian in the Hilbert space $\mathcal{H}$ to an effective time-independent Hamiltonian $H_{\text{eff}}$ such that the unitary evolution coincides with $e^{iH_{\text{eff}}\,t}$ when $t = nT$. In order to achieve this, we conceive a family of Hamiltonians $H(l;t)$, parametrized by $l$, all of which are linked to each other by unitary transformations given by the operation $P_{l_1,l_2;t}$, i.e.

$$H(l';t) = P_{l',l;t}[H(l;t)] \equiv Z^{\dagger}(l',l;t)H(l;t)Z(l',l;t) - i\,Z^{\dagger}(l',l;t)\left\{\frac{\partial}{\partial t}Z(l',l;t)\right\} \forall\, l,l', \tag{S8}$$

and $H(0;t)$ is our original Hamiltonian $H(t)$. In addition, we conceive a parameter value $\bar{l}$ such that $H_{\text{eff}} \approx H(\bar{l};t) = P_{\bar{l},0;t}[H(t)]$ is approximately time independent. This leads to the relation

$$U(t) \approx e^{-iH_{\text{eff}}t}Z(t) \tag{S9}$$

where we have dropped the dependencies on $\bar{l}, l$, and $U(t)$ solves the Schrödinger equation for the original Hamiltonian viz. $i\partial_t U(t) = H(t)U(t)$. Note that Floquet's theorem ensures (see [r4, r5] and references therein) that $Z(t)$ is time-periodic if $H(\bar{l};t)$ is time independent. If the family of transformations can be generated in such a way that $U(\bar{l},l;0) = \mathbb{1}$, then this result can be combined with Eq. S9 to yield $U(t) \approx e^{-iH_{\text{eff}}t}$ at integer multiples of the period. Thus, the claim made in the main paper that $H_{\text{eff}}$ is *dynamically equivalent* to $H(t)$ when strobed at multiples of the time period is verified. The Wilson (Wegner) numerical renormalization group provides a systematic approach towards determining such a family of transformations, as well as the value $\bar{l}$, via unitary flows [r6]. The three step prescription restricts such flows to the Floquet-Hilbert space that is appropriate for time-periodic Hamiltonians, and is detailed below.

The first of the three-step process, rewriting the spin Hamiltonian in bosonic language, has already been performed in Eq. S1. In the next step, we separate the Hamiltonian in Eq. S1 into $H(t) = H_s + H_d(t)$, where

$$H_s(J_i, h_i) = -\alpha J \sum_i^{L-1} J_i \left(c_i^{\dagger}c_{i+1}^{\dagger} + c_i^{\dagger}c_{i+1} + \text{h.c.}\right) - 2\alpha \sum_i^L h_i c_i^{\dagger}c_i ,$$

$$H_d(J_i, t) = -2\,h(t) \sum_i^{L-1} c_i^{\dagger}c_i , \tag{S10}$$

This Hamiltonian is defined in a complete Hilbert space $\mathcal{H}$. In this Hilbert space, we transform the Hamiltonian by the unitary transformation $\Lambda(t) = \exp\left\{i\int_0^t d\tau\, H_d(\tau)\right\}$, yielding

$$\tilde{H}(t) = \Lambda(t)\,H(t)\,\Lambda^{\dagger}(t) - i\,\Lambda(t)\,\left\{\frac{\partial}{\partial t}\Lambda(t)\right\}^{\dagger}$$

$$= \Lambda(t)\,H_s\,\Lambda^{\dagger}(t). \tag{S11}$$

This Hamiltonian, like the previous one, is periodic in time with period $T$. In that case, Floquet's theorem states that the corresponding Schrödinger equation can be solved by states of the type $|\psi_p(t)\rangle = e^{i\Omega_p t}\,|\phi_p(t)\rangle$, where $|\phi_p(t+nT)\rangle = |\phi_p(t)\rangle$, and the Floquet quasienergies $\Omega_p$ are real. Using Fourier's theorem, the Floquet eigenstates $|\phi_p(t)\rangle$ can be expanded in a basis $|n\rangle$, where $\langle t|n\rangle = e^{-i\,n\omega t}$. These basis states span a Hilbert space $\mathcal{H}_T \subset \mathcal{H}$, consisting of states that are time-periodic in $T$, yielding $\tilde{H}(t) = \sum_n \tilde{H}_n\,e^{-i\,n\omega t}$, with time-independent operators $\tilde{H}_n$. The corresponding Floquet operator $K(t) \equiv \tilde{H}(t) - i\,\partial_t$ is now mapped to an analogous operator $\mathcal{K}$ in the Floquet - Hilbert space $\mathcal{H} \otimes \mathcal{H}_T$ to yield

$$\mathcal{K} = \tilde{H}_0 \otimes \mathbb{1} + \mathbb{1} \otimes \omega\,\hat{n} + \sum_{n \neq 0} \tilde{H}_n \otimes \sigma_n. \tag{S12}$$

In Eq. S12, the operator $i\,\partial_t$ has been mapped to $\mathbb{1} \otimes \omega\,\hat{n}$. Here, the Floquet number operator acts only in $\mathcal{H}_T$ via the basis states as $\hat{n}\,|n\rangle = n\,|n\rangle$. In addition, the operators $\sigma_n \in \mathcal{H}_T$ act as ladder operators on the basis states with $\sigma_m\,|n\rangle = |n+m\rangle$. Note that $\sigma_0 = \mathbb{1}$, and the $n = 0$ term in the structure of $\mathcal{K}$ is shown separately in Eq. S12. Starting from Eq.s S11 and S10, as well as the Fourier expansion above, we can obtain the components $\tilde{H}_n$. Using the Jacobi-Anger expansion [r7], $\exp\left[i\,\eta\,\sin\omega t\right] = \sum_m \mathcal{J}_m(\eta)\,e^{-in\omega t}$, we obtain

$$\tilde{H}_0 = -\alpha J \sum_{i=1}^{L-1} \mathcal{J}_0(\eta) J_i \left(c_i^{\dagger}c_{i+1}^{\dagger} + c_i^{\dagger}c_{i+1} + \text{h.c}\right) - 2\,\alpha \sum_{i=1}^{L-1} h_i c_i^{\dagger}c_i, \tag{S13}$$

with $\eta \equiv 4\, h_0/\omega$, and the $n \neq 0$ components given by $\tilde{H}_n = \sum_i \tilde{H}_n^{(i)}$, where

$$\tilde{H}_n^{(i)} \equiv -\alpha J J_i \left\{ \mathcal{J}_n(\eta) \left( c_i^\dagger c_{i+1}^\dagger + c_i^\dagger c_{i+1} \right) + \mathcal{J}_{-n}(\eta) \left( c_{i+1} c_i + c_{i+1}^\dagger c_i \right) \right\}. \tag{S14}$$

The ultimate purpose of writing our Hamiltonian in this manner is to use a systematic approach to renormalize the hopping terms in $\tilde{H}_n^{(i)}$ to determine any regime in the parameter space where they all might be ignored. The final step *viz.* the execution of this approach is described below.

The final step involves determining unitary transformations in $\mathcal{K}$ that lead to a dynamically equivalent expression whose time dependent parts ($\tilde{H}_{n \neq 0}$) vanish, yielding an effective Hamiltonian $H_{\text{eff}} = \tilde{H}_0$. In order to do so, we insert an unknown parameter $l$ and seek a family of unitary transformations on $\mathcal{K}(l) \in \mathcal{H} \otimes \mathcal{H}_T$ generated by choosing antihermitian operators $\Gamma(l) \in \mathcal{H} \otimes \mathcal{H}_T$, such that $\mathcal{K} = \mathcal{K}(l)|_{l=0}$, and the flow equations,

$$\frac{\mathrm{d}}{\mathrm{d}l} \mathcal{K}(l) = \Big[ \Gamma(l),\, \mathcal{K}(l) \Big], \tag{S15}$$

are satisfied by all $\mathcal{K}(l)$ while maintaining the same component structure as shown in Eq. S12. This way, our theory remains 'renormalizable' for all $l$. Any terms that do not belong in one of the components are 'non-renormalizable', and are reduced iteratively by generating new trials for $\Gamma(l)$ that absorb these terms. The fixed point of Eq. S15 (at, say $l = \bar{l}$) should yield the time-independent Hamiltonian via the solution $U_s^\dagger(\bar{l}, t) H_{\text{eff}} U_s(\bar{l}, t)$, where $U_s(l, t)$ are the unitary transformations that link the different $\mathcal{K}(l)$ to $\mathcal{K}(0)$. In order to obtain the first iterative solution for Eq. S15, we choose for $\Gamma(l)$ the Wilson generator [r8–r10]

$$\Gamma_1(l) = \sum_{n \neq 0} b_n(l) \left[ \mathbb{1} \otimes \omega \hat{n}, \tilde{H}_n \otimes (\sigma_n - \mathbb{1}) \right], \tag{S16}$$

and the trial solution for the corresponding flow in Eq. S15,

$$\mathcal{K}_1(l) = \tilde{H}_0 \otimes \mathbb{1} + \mathbb{1} \otimes \omega\, \hat{n} + \sum_{j, m \neq 0} b_m(l)\, \tilde{H}_m(J_j) \otimes \sigma_m, \tag{S17}$$

with the initial condition $b_m(0) = 1$. In Eq. S16, the RHS is the commutator between the only term in $\mathcal{K}(l)$ that is fully diagonal in $\mathcal{H} \otimes \mathcal{H}_T$ [r9], and the time-dependent parts of $\mathcal{K}(l)$ that come from the nonzero Fourier modes of $\tilde{H}(t)$. The choice of $(\sigma_n - \mathbb{1})$ over $\sigma_n$ in Eq. S16 ensures that the unitary transformation that takes $H(t)$ to $H_{\text{eff}}$ is unity at $t = 0$ [r10]. Substituting Eqs. S17 and S16 into Eq. S15 yields a dynamical system after comparing terms on both sides, $b_n'(l) = -\omega^2 n^2 b_n(l)$, or $b_n(l) = \exp\left(-n^2 \omega^2 l\right)$. Here, we have dropped all non-renormalizable terms in the RHS of Eq. S15. These are now sent to the next iteration, which yields the dominant contribution in $\omega$. The new generator, chosen in a manner similar to the previous iteration, is

$$\Gamma_2(l) = \Gamma_1(l) + 2\alpha^2 J \sum_{i, n \neq 0} b_n(l)\, J_i\, h_i \left\{ \mathcal{J}_n(\eta) \hat{f}_i^\dagger - \mathcal{J}_{-n}(\eta) \hat{f}_i \right\} \otimes (\sigma_n - \mathbb{1}),$$

$$\hat{f}_i^\dagger \equiv c_{i-1}^\dagger \left( c_i - c_i^\dagger \right) - c_i^\dagger \left( c_{i+1}^\dagger + c_{i+1} \right). \tag{S18}$$

Analogously, the trial for the next iteration is

$$\mathcal{K}_2(l) = \mathcal{K}_1(l) + \alpha \sum_i J_i\, h_i \left\{ a_+(l) \hat{f}_i^\dagger - a_-(l) \hat{f}_i \right\} \otimes \mathbb{1}, \tag{S19}$$

with new parameters $a_\pm(l)$, and the initial conditions $a_\pm(0) = 0$, $b_m(0) = 1$ keep $\mathcal{K}(0) = \mathcal{K}$. Substituting these into Eq. S15 and comparing like terms on both sides yields the following nontrivial flow dynamics for the parameters,

$$\frac{\mathrm{d}b_n}{\mathrm{d}l} = -\omega^2 n^2 b_n,$$

$$\frac{\mathrm{d}a_+}{\mathrm{d}l} = -2\alpha J \omega \sum_{n \neq 0} n b_n(l)\, \mathcal{J}_n(\eta),$$

$$\frac{\mathrm{d}a_-}{\mathrm{d}l} = +2\alpha J \omega \sum_{n \neq 0} n b_{-n}(l)\, \mathcal{J}_n(\eta), \tag{S20}$$

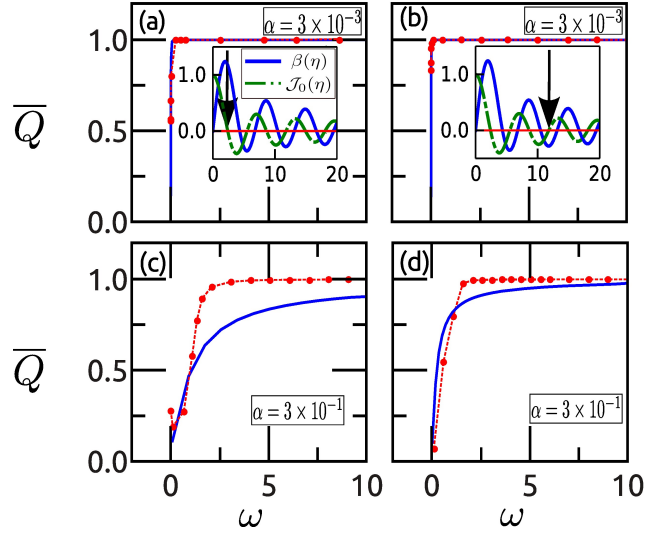

FIG. S2. (Color Online) Plots of $\overline{Q}$ v.s. $\omega$ for two different values of $\alpha$. In the main panels, the red circles (blue curves) plot the ordinate obtained from the time-evolution of the full Hamiltonian given by Eq. S1 (effective Hamiltonian given by Eqs. S23) averaged over $t = [0, 300]$, and $10^3$ disorder realizations. The values of $\alpha$ chosen are indicated in the main legends. Panels **(a)** and **(c)**: The value of $\eta = 4h_0/\omega$ was kept fixed at the first root of $\mathcal{J}_0(\eta)$ as indicated by a gray arrow next to the plot of $\mathcal{J}_0$ (green dashed) in the inset on panel (a). Panels **(b)** and **(d)**: Here, $\eta$ was fixed at the fourth root as indicated in the inset on panel (b). In both insets, the blue curves plot $\beta(\eta)$.

where we have truncated out non-renormalizable terms [r11] by substituting $a_\pm(l) = a_\pm(0) = 0$ on the RHS of the flow equation for this iteration [r10]. These equations yield the same solution for $b_n$ as the previous iteration, and

$$a_\pm(l) = \mp \frac{2\alpha J}{\omega} \sum_{n \neq 0} \left(1 - e^{-n^2 \omega^2 l}\right) \frac{\mathcal{J}_n(\eta)}{n}. \tag{S21}$$

The fixed point of the flow dynamics occurs when $l \to \infty$, where $a_\pm(\infty) = \mp 2\alpha J \beta(\eta)/\omega$, and $\beta(\eta) \equiv \sum_{n \neq 0} \mathcal{J}_n(\eta)/n$. Substituting these in Eqs. S19 and S17 yields

$$\mathcal{K}_2(\infty) = \left\{ \tilde{H}_0 - \frac{2\alpha^2 J}{\omega} \beta(\eta) \sum_i J_i h_i \left(\hat{f}_i + \hat{f}_i^\dagger\right) \right\} \otimes \mathbb{1} + \mathbb{1} \otimes \omega \hat{n} + \mathcal{O}(\alpha/\omega^2). \tag{S22}$$

In Eq. S22, we identify the first part in the sum on the RHS with the time-independent part of the Hamiltonian, $\tilde{H}_0 \in \mathcal{H}$, the second part with the operator $-i\,\partial_t \in \mathcal{H}_T$, and all the $\tilde{H}_{n \neq 0}$ as the time-dependent parts, which scale as $\mathcal{O}(\omega^{-2})$. Thus, this new Floquet operator maps to a time-independent Hamiltonian $H_{\text{eff}} \in \mathcal{H}$ up to $\mathcal{O}(\omega^{-1})$, viz.

$$\begin{aligned}
H_{\text{eff}} &= -J \sum_i j_i^{(0)} \left(c_i^\dagger c_{i+1}^\dagger + \text{h.c.}\right) - J \sum_i j_i^{(1)} \left(c_i^\dagger c_{i+1} + \text{h.c}\right) - \mu \sum_i j_i^{(2)} c_i^\dagger c_i, \\
j_i^{(0)} &\equiv \alpha J_i \left\{ \mathcal{J}_0(\eta) - \frac{4\alpha h_i}{\omega} \beta(\eta) \right\}, \\
j_i^{(1)} &\equiv \alpha J_i \, \mathcal{J}_0(\eta), \\
j_i^{(2)} &\equiv \frac{h_i}{J}, \\
\mu &\equiv 2\alpha J.
\end{aligned} \tag{S23}$$

Note that, in order to obtain $H_{\text{eff}}$, we have canceled out terms like $\sum_i c_{i-1}^\dagger c_i$ with terms like $\sum_i c_i^\dagger c_{i+1}$ while summing over large number of lattice indices. Setting the field disorder $h_i$ to zero recovers the analytical results discussed in the main paper, where the full Hamiltonian maps via unitary flows to an *identical* effective Hamiltonian with renormalized hopping $J \to J \mathcal{J}_0(\eta)$.

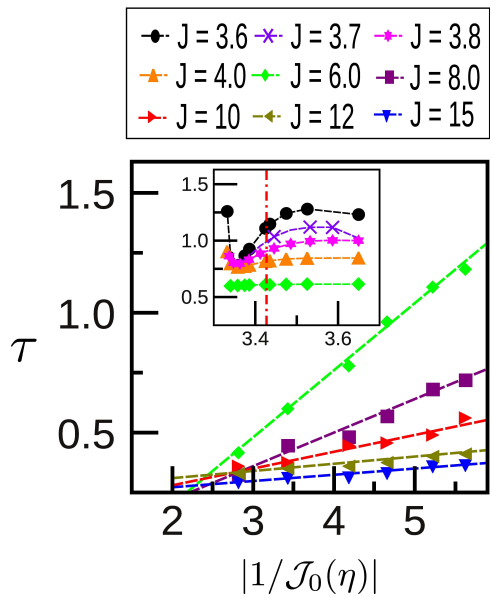

FIG. S3. (Color Online) Here we plot $\tau$ vs $1/\mathcal{J}_0(\eta)$ for discrete values of $\eta$ for which $\beta(\eta) = 0$ (six smallest roots) is satisfied. Plots for different values of $J$ are shown. We observe a linear behavior (discrete data points fitted with least-square lines) in the regime $\mu \ll J$. This can be explained from the fact that for $\beta(\eta) = 0$ and $\mu \ll J$, $H_{eff} \propto \mathcal{J}_0(\eta)$ (Eq. S23) to a good approximation, and hence all timescales (including) $\tau$ should be proportional to $1/\mathcal{J}_0(\eta)$. The inset shows continuous plots of $\tau$ vs $1/|\mathcal{J}_0(\eta)|$ in the immediate neighborhood of a root of $\beta(\eta)$. The root is indicated by a vertical (dot-dashed) red line, and the values of $J$ chosen are indicated in the main legend.

## COMPARISON WITH NUMERICS

The validity and limits of our theoretical description in the previous section can be quantitatively determined by evolving the dynamics in Eq. S5 in the first section with the time-independent $H_{\text{eff}}$ in Eq. S23. Switching to the fermion representation leaves the $H_{\text{eff}}$ invariant, and the matrices $A(t), B(t)$ are time independent. The nonzero matrix elements in the dynamics for $H_{\text{eff}}$ are given by $A_{ij} = -\alpha h_i\,\delta_{ij}$, $B_{i,i+1} = -B_{i+1,i} = \alpha J j_i/2$ at any freezing point given by $\mathcal{J}_0(\eta) = 0$. The quantities $\overline{Q}$ and $\overline{m^z(t)}$ can be extracted and compared with those from the evolution of Eq. S5 with the full Hamiltonian in Eq. S1. Figure S2 shows plots comparing the approach to freezing in $\overline{Q}$ with increasing $\omega$ for both the time evolution of $H(t)$ and $H_{\text{eff}}$. The left panels show plots when $h_0$ is adjusted for each $\omega$ such that $\eta$ is at the first root of the zeroth order Bessel function, and the right panels show similar plots for the fourth root. Based on the structure of our truncated approximation for $H_{\text{eff}}$ in Eq. S23, we expect that higher order contributions will be controlled by powers of the disorder strength $\alpha$, as well as the value of $\beta(\eta)$ at freezing. This is borne out by the data shown in Fig. S2. When $\eta$ is at the first root of $\mathcal{J}_0$, the value of $\beta(\eta)$ is near-maximum (see the insets of Fig. S2). Here, contributions from the higher order terms cause theory to differ from the numerics quantitatively if $\alpha$ is appreciable. This can be seen in Fig S2 (c), where $\alpha = 0.3$. However, these deviations are smaller when $\alpha$ is decreased to 0.003 as seen in Fig S2 (a). The function $\beta(\eta)$ decreases non-monotonically with increasing $\eta$ (insets in Fig S2), suppressing higher order contributions at larger roots. Note that $\mathcal{J}_0(\eta)$ and $\beta(\eta)$ never vanish simultaneously. The best agreement is in Fig S2 (b), where $\alpha$ and $\beta(\eta)$ are the smallest among the chosen values at freezing.

Another aspect of the exact dynamics that can be inferred from the effective Hamiltonian in Eq. S23 is the time-scale of the magnetization at or near the roots of the function $\beta(\eta)$. If we switch off the field disorder $h_i$ by replacing them with a constant transverse field $h_i = J$, then, in the immediate neighborhood of a root of $\beta(\eta_r)$, a Taylor expansion yields $j_i^{(0)} \approx \alpha J_i\{\mathcal{J}_0(\eta_r) - (4\alpha J/\omega)(\eta - \eta_r)\beta'(\eta_r)\}$, where $\beta'$ denotes the $\eta-$ derivative of $\beta$. Since $\eta$ and $\eta_r$ both scale inversely with $\omega$, varying $\eta$ by varying $h_0$ alone leads to a $\omega^{-2}-$ dependence on the second term, allowing us to ignore it for large $\omega$. Thus, the contribution of the transverse field at or near the roots of $\beta$ is negligible in this regime. In this regime, the time-scale $\tau$ of the magnetization $m^z(t)$ is expected to scale linearly with the inverse of the renormalized hopping $viz.$ $1/|J\mathcal{J}_0(\eta)|$ provided $J$ is sufficiently large compared to $\alpha$ so as to minimize the dynamical effects of the bond disorder. Figure S3 shows plots of the time-scale $\tau$, obtained via exponential fit to the decay of $m^z(t)$, at the first six roots of $\beta(\eta)$. The data is plotted against $1/|\mathcal{J}_0(\eta)|$, and the points fitted by least-squares to

lines. The plots demonstrate that the fits get progressively better as $J$ is increased, with deviations due to bond disorder becoming more and more prominent at smaller $J$. The inset also shows continuous plots of $\tau$ vs. $1/|\mathcal{J}_0(\eta)|$ in the immediate neighborhood of a single root of $\beta(\eta)$ (shown as a vertical red line), demonstrating approximately linear behavior in the vicinity of the root. Deviations from linearity occur for larger values of $\eta$, where higher order terms in the Taylor expansion begin to contribute. These contributions are non-monotonic, however, and reduce as $\eta$ approaches another root of $\beta$.



\end{document}